# Nonlinear Susceptibilities of a Weakly Disordered Uniaxial Ferromagnet in the Critical Region

D. V. Pakhnin[1]*, A. I. Sokolov[1], and B. N. Shalaev[2]

[1] *St. Petersburg State Electrotechnical University (LÉTI), St. Petersburg, 197376 Russia*
[2] *Ioffe Physicotechnical Institute, Russian Academy of Sciences, St. Petersburg, 194021 Russia*
*\*e-mail: ais@sokol.usr.etu.spb.ru*

For a weakly disordered three-dimensional Ising model in the critical region, the sixth-order effective coupling constant and the fourth- and sixth-order nonlinear susceptibilities are determined. The values of these quantities are found to differ radically (by a factor of 1.5–3) from the corresponding values in a pure ferromagnet, and their measurement is suggested for the identification of the critical behavior of impure systems.

From the mid-1970s, the critical thermodynamics of three-dimensional impurity systems has been the object of intensive studies, both theoretical and experimental. Theoretical achievements, such as the determination of the mechanism governing the effect of impurities on the critical behavior, the formulation of the Harris criterion, the construction of the $\sqrt{\epsilon}$ expansion, and the calculation of the critical indices and critical amplitude ratios in the framework of the perturbation theory [1–13], have stimulated subsequent studies, the development of which in the last few years acquired the character of an explosion. Advancement in this field of research was, to some extent, caused by the discovery of the fact that, for the systems under discussion, an increase in the order of the renormalized perturbation theory does not lead to stabilization of the numerical results for the critical indices and other universal physical quantities. This feature is in contradiction with the known properties of renormalized group expansions for pure systems, which allow one, by applying the appropriate resummation procedures, to determine the universal parameters with an accuracy progressively increasing from order to order [14–22]. Most likely, the aforementioned anomaly, which manifests itself only in the five-loop and six-loop approximations [23–26], reflects the much discussed Borel nonsummability of renormalization group expansions for impurity systems (see, e.g., [27–29] and recent reviews [30–32]).

The absence of convergence of the iteration procedures based on the renormalization group theory of perturbations does not, however, preclude one from obtaining numerical estimates of the critical indices with an acceptable accuracy. The latter implies a relatively small scatter of the results obtained from different approximations, the insensitivity of the results to changes in the resummation technique, and, evidently, a good agreement between the theoretical predictions and the results of physical and computer experiments. For example, for the critical index of susceptibility $\gamma$ of the impurity three-dimensional Ising model, the four-, five-, and six-loop approximations yield the values 1.326–1.321 [10, 11], 1.325 [25], and 1.330 [26], respectively, and the variations of $\gamma$ in passing from one resummation technique to another do not exceed 0.01. This suggests that the field-theoretical renormalization group method can be used for calculating other universal critical parameters of three-dimensional impurity systems.

Below, we determine the nonlinear susceptibilities of the fourth ($\chi_4$) and sixth ($\chi_6$) orders and the effective coupling constant $v_6$ for a weakly disordered three-dimensional Ising ferromagnet in the critical region. At $T \longrightarrow T_c$, these quantities, just like the linear susceptibility $\chi$ and other equilibrium parameters, take on universal asymptotic values, which can be measured with high accuracy in modern experiment.

The free energy of a uniaxial ferromagnet as a function of magnetization $M$ in an external magnetic field $H$ can be represented in the form

$$F(M, m) = F(0, m) + \frac{1}{2}m^{2-\eta}M^2 + m^{1-2\eta}v_4 M^4 + m^{-3\eta}v_6 M^6 + \ldots - HM, \quad (1)$$

where $m$ is the inverse correlation radius, $\eta$ is the Fisher index, and $v_4$ and $v_6$ are the effective coupling constants taking universal critical values at the Curie point.

Using expansion (1), one can easily express the nonlinear susceptibilities $\chi_4$ and $\chi_6$ in terms of $\chi$, $v_4$, and $v_6$:

$$\chi_4 = \left.\frac{\partial^3 M}{\partial H^3}\right|_{H=0} = -24\xi^2 m^{-3} v_4,$$

$$\chi_6 = \left.\frac{\partial^5 M}{\partial H^5}\right|_{H=0} = 720\chi^3 m^{-6}(8v_4^2 - v_6). \quad (2)$$

Thus, the determination of the nonlinear susceptibilities in the critical region is reduced to the calculation of the universal asymptotic values of $v_4$ and $v_6$.

For weakly disordered systems, the thermodynamic quantities are determined by averaging over random impurity configurations. This averaging is most simply performed by the replica technique. The latter is based on the fluctuation Hamiltonian of the $n$-vector cubic model

$$H = \int d^3x \left[\frac{m_0^2 \varphi_\alpha^2 + (\nabla \varphi_\alpha)^2}{2} + u_4^{(0)} \varphi_\alpha^2 \varphi_\beta^2 + v_4^{(0)} \varphi_\alpha^4\right], \quad (3)$$

which, in the limit $n \longrightarrow 0$, reproduces the critical behavior of the impurity Ising model. This behavior is controlled by the fixed impurity point of the renormalization group equations. The point in question is a stable node in the $(u_4, v_4)$ plane, and its coordinates are known in the highest approximation available, i.e., in the six-loop approximation [26]. It is essential that the case of a nonzero external magnetic field corresponds to the $n$-component cubic model in a uniform field directed along the principal diagonal of the hypercube. It can be shown that, for the solution that does not violate the replica symmetry, the coupling constant $u_4$ drops out of the equation of state in the limit $n \longrightarrow 0$. From the physical point of view, this is quite important, because, in this case, the "wrong" sign of $u_4$ does not lead to the instability of the effective Hamiltonian.

Thus, at the Curie point, the effective fourth-order coupling constant in Eq. (1) is equal to the coordinate $v_4^*$ of the fixed impurity point. Hence, the asymptotic behavior of $\chi_4$ at $T \longrightarrow T_c$ is determined by the quantity $v_4^*$. The situation with the nonlinear susceptibility $\chi_6$ is more complicated. The determination of its critical asymptotic behavior involves the calculation of the effective coupling constants $u_6$, $q_6$, and $v_6$ for the model given by Eq. (3). These constants act as coefficients of the invariants $M_\alpha^2 M_\beta^2 M_\gamma^2$, $M_\alpha^4 M_\beta^2$, and $M_\alpha^6$ in the expansion of the free energy of the cubic model. Recently, the quantities $u_6$, $q_6$, and $v_6$ were determined in the form of renormalized perturbative series expansions in the four-loop approximation [33]. For the $O(n)$-symmetric systems, the series of length as large as this allow one to calculate the universal values of the sixth-order coupling constant with an accuracy no lower than 1% [19, 22]. Since the expression for $\chi_6$ involves only one of the three coupling constants, namely, $v_6$, we present the renormalization group expansion only for this constant. In the limit $n \longrightarrow 0$, the expansion has the form

$$\frac{v_6}{v_4^2} = \frac{9}{\pi}(2u_4 + v_4 - 2.9001567 u_4^2 - 3.1830989 u_4 v_4$$

$$- 0.9549296 v_4^2 + 5.579725 u_4^3$$

$$+ 10.03487 u_4^2 v_4 + 6.222000 u_4 v_4^2 \quad (4)$$

$$+ 1.389963 v_4^3 - 12.5233 u_4^4$$

$$- 31.7631 u_4^3 v_4 - 30.6484 u_4^2 v_4^2$$

$$- 13.8874 u_4 v_4^3 - 2.50173 v_4^4).$$

Note that the quantities $u_4$ and $v_4$ differ by a factor of $\pi/4$ from their analogs $u$ and $v$ used in [33].

Series (4) is of the asymptotic type. However, series of this kind allow one to obtain reliable quantitative results by applying the appropriate resummation techniques. One of them is used in our calculations in [11]. At the first step, expansion (4) is transformed to a convergent series with the help of the Borel–Leroy generalized transformation

$$f(u, v) = \sum_{ij} c_{ij} u^i v^j = \int_0^\infty e^{-t} t^b F(ut, vt) dt,$$

$$F(x, y) = \sum_{ij} \frac{c_{ij} x^i y^j}{(i + j + b)!}. \quad (5)$$

Then, using the Borel transform of the initial function, we construct an auxiliary series

$$\tilde{F}(x, y, \lambda) = \sum_{n=0}^\infty \lambda^n \sum_{l=0}^n \frac{c_{l, n-l} x^l y^{n-l}}{n!}, \quad (6)$$

with the coefficients as homogeneous polynomials in the variables $u_4$ and $v_4$. To perform analytic continuation beyond the circle of convergence, Padé approximants $[L/M]$ in the variable $\lambda$ are used, the value of this variable being set equal to unity at the terminal step. The described resummation procedure retains all point symmetry properties of the initial expansions [34] and provides rapid convergence of the iteration process if the Borel summability of the renormalized group series takes place.

Since the parenthetical expression in Eq. (4) is a fourth-order polynomial, we can construct four different Padé approximants: [3/1], [2/2], [1/3], and [0/4]. It is well known that the diagonal approximants $(L = M)$ or approximants close to them possess the best approximating properties. However, with an increase in the denominator exponent $M$, the number of the approximant poles in the complex plane also increases, and when these poles fall on the positive real semiaxis or close to it, they can make the approximant unsuitable for series summation. Therefore, in the resummation of expansion (4), we use only approximants [3/1] and [2/2]. In addition, our analysis also includes approximant [2/1], which practically corresponds to the use of the three-loop approximation. The above operations are directed toward the aim of revealing the sensitivity of the numerical results to the approximation order and to obtain additional information for the optimization of the resummation procedure by choosing the optimal value of the free parameter b in the Borel–Leroy transformation.

**Table**

| b | | 0 | 1 | 2 | 3 | 5 | 10 | 15 | 20 |
|---|---|---|---|---|---|---|---|---|---|
| $u_4^* = -0.50$, | [2/2] | 2.056 | – | – | – | 2.161 | 2.120 | 2.109 | 2.103 |
| $u_4^* = 1.53$ | [3/1] | 2.319 | 2.255 | 2.216 | 2.190 | 2.156 | 2.117 | 2.100 | 2.090 |
| (6-loop) | [2/1] | 1.960 | 2.033 | 2.072 | 2.095 | 2.123 | 2.153 | 2.165 | 2.172 |
| $u_4^* = -0.53$, | [2/2] | 2.062 | – | – | – | – | 2.150 | 2.135 | 2.127 |
| $u_4^* = 1.57$ | [3/1] | 2.364 | 2.296 | 2.255 | 2.226 | 2.191 | 2.149 | 2.131 | 2.120 |
| (6-loop) | [2/1] | 1.957 | 2.034 | 2.074 | 2.099 | 2.129 | 2.160 | 2.172 | 2.180 |
| $u_4^* = -0.56$, | [2/2] | 1.867 | – | – | – | – | 1.995 | 1.973 | 1.963 |
| $u_4^* = 1.58$ | [3/1] | 2.188 | 2.125 | 2.087 | 2.061 | 2.028 | 1.990 | 1.973 | 1.963 |
| (5-loop) | [2/1] | 1.762 | 1.834 | 1.871 | 1.895 | 1.922 | 1.951 | 1.963 | 1.969 |

The results of our calculations are shown in the table, which presents the effective coupling constant $v_6$ as a function of the parameter b. The values of $v_6$ are determined at the fixed impurity point using the three aforementioned Padé approximants. Since the coordinates of the fixed impurity point, $u_4^*$ and $v_4^*$, are known with a limited accuracy, we calculated the universal critical value of $v_6$ for three sets of $u_4^*$ and $v_4^*$. The first two of them ($u_4^* = -0.50$, $v_4^* = 1.53$ and $u_4^* = -0.53$, $v_4^* = 1.57$) are obtained in the six-loop approximation with the use of two different resummation strategies [26], and the third set ($u_4^* = -0.56$, $v_4^* = 1.58$) is determined from the five-loop expansions subjected to the resummation by the Padé–Borel–Leroy method [25]. The empty cells in the table mean that, for the corresponding values of b, the Padé approximant [2/2] has "dangerous" poles.

As one can see from the table, the numerical values of $v_6$ obtained with the three chosen Padé approximants weakly depend on the parameter b, and, for each set of $u_4^*$ and $v_4^*$, one can easily determine the optimal value of b at which the three approximants yield coincident or very close results. This fact points to the high efficiency of the resummation technique used in our calculations. The analysis of the data presented in the table shows that three variants of the coordinates taken for the fixed impurity point correspond to the estimates $v_6 = 2.14$, 2.15, and 1.96, respectively. Since the coordinates $u_4^*$, $v_4^*$ determined from the six-loop approximation should be considered as the most reliable ones and the processing of the divergent series (4) can hardly provide an accuracy better than to the second decimal place, we accept the following final result of our calculations:

$$v_6 = 2.1 \pm 0.2. \tag{7}$$

The chosen error limits are rather conservative, and, hence, the true asymptotic value of $v_6$ is certain to lie within the interval bounded by Eq. (7).

It is of interest to compare the universal critical value of $v_6$ for the impurity Ising model with its analog for a pure (defect-free) system. The factor that really characterizes the contribution of the effective coupling constant $v_6$ to the equation of state is the ratio $v_6/v_4^2$. Taking the average value $v_4^* = 1.55$ as the coordinate of the fixed impurity point, we obtain the following ratio for the disordered Ising model at the critical point: $v_6/v_4^2 = 0.87$. For a pure uniaxial ferromagnet, the corresponding ratio is $v_6/v_4^2 = 1.64$–1.65 [17, 18, 35–38]. Thus, the impurities reduce this ratio almost by half. Since the ratio under discussion appears in the equation of state and, hence, is available for experimental study, the measurement of $v_6/v_4^2$ can be used to identify the critical behavior of impurity systems.

An equally great difference is observed between the nonlinear susceptibilities of impurity and pure Ising ferromagnets. For the three-dimensional Ising model, we have $v_4^* = 0.99$ [14, 15, 17, 18]. Then, according to the first expression in Eqs. (2), the value of the universal combination $\chi_4\chi^{-2}m^3$ calculated for an impurity ferromagnet is 55–60% greater than for a pure ferromagnet. The difference in the sixth-order nonlinear susceptibilities is even more substantial. According to Eqs. (2), for a pure uniaxial ferromagnet, we have $\chi_6\chi^{-3}m^6 = 4.5 \times 10^3$, whereas for a weakly disordered system, we have $\chi_6\chi^{-3}m^6 = 12.3 \times 10^3$. This almost threefold change in the parameter $\chi_6\chi^{-3}m^6$ under the effect of impurities can certainly be detected experimentally.

This work was supported by the Russian Foundation for Basic Research (project nos. 01-02-17048, 01-02-17794) and the Ministry of Education of the Russian Federation (project no. E00-3.2-132). A.I. Sokolov and D.V. Pakhnin are also grateful to the International Science Foundation and the St. Petersburg City Administration for financial support in the framework of individual projects (nos. p2001-90 and s2001-1002).